\documentclass[reprint,longbibliography,
 amsmath,amssymb,
 prd,superscriptaddress]{revtex4-1}
\usepackage{graphicx}
\usepackage{dcolumn}
\usepackage{bm}
\usepackage[colorlinks=true, citecolor=blue, linkcolor=blue, urlcolor=blue]{hyperref}
\usepackage{xcolor}
\usepackage{amsmath, amssymb, amsfonts}
\usepackage{mathtools}
\usepackage{subfigure}
\usepackage{mathrsfs}
\usepackage{bbold}
\usepackage{sidecap}
\usepackage{verbatim}

\definecolor{purple1}{rgb}{128,0,128}

\newcommand{\nn}{\nonumber\\}
\usepackage{bigints}
\newcommand{\bea}{\begin{eqnarray}}
\newcommand{\ea}{\end{eqnarray}}

\definecolor{darkpastelgreen}{rgb}{0.01, 0.75, 0.24}
\begin{document}
\title{Fluid-dynamical analogue of nonlinear gravitational wave memory} 
\title{Inherent nonlinearity of fluid motion and acoustic gravitational wave memory}
\author{Satadal Datta} 
\author{Uwe R. Fischer}
\affiliation{Seoul National University, Department of Physics and Astronomy, Center for Theoretical Physics, Seoul 08826, Korea}
\date{\today}

\begin{abstract}
{We consider the propagation of nonlinear sound waves in a 
perfect fluid at rest. By employing the Riemann wave equation of nonlinear acoustics in one spatial dimension,  
it is shown that waves carrying a constant density perturbation at their tails produce an acoustic analogue of 
gravitational wave memory. For the acoustic memory, which is in general {\em nonlinear}, 
the nonlinearity of the effective spacetime dynamics is not due to the 
Einstein equations, but due to the nonlinearity of the perfect fluid equations. For concreteness,  we employ a box-trapped Bose-Einstein condensate, and suggest an experimental protocol to observe acoustic gravitational wave memory.  }
\end{abstract}

\maketitle
\section{Introduction} 
The pinnacle in the history of 21st century experimental physics so far was undoubtedly 
the first detection of gravitational waves (GWs) \cite{abbott2016observation}. 
One of the most remarkable features of GWs
is that they can present {\em memory} \cite{Zeldovich,disturbing,BT_Nature}. 
A freely falling test mass (representing an ideal detector) 
is permanently displaced after the GW train has passed.
The memory effect is present already in 
linearized Einstein theory.   
There exists however also a genuinely {\em nonlinear} GW memory which is 
arising from the nonlinearity of the Einstein equations in the GW amplitude \cite{christodoulou1991nonlinear}. 
A physically particularly lucid explanation of the nonlinear GW memory effect, 
derived in a veritable mathematical tour de force by Christodoulou, was given by Thorne shortly thereafter, and was providing an interpretation of nonlinear memory in terms of the gravitons emitted by the GW source \cite{thorne1992gravitational}. 

{Proposals to detect the GW memory effect have been put forward, e.g., in 
\cite{favata2010gravitational, johnson2019prospects,Boersma}, where the Christodoulou
nonlinear memory has already argued early on to be even much more feeble than its linear counterpart \cite{Will,johnson2019prospects}. 
The permanent displacement of test masses after the wave train has passed 
is generally far more difficult to detect than the oscillatory motion  induced  by the GW pulse.
GW memory, both in its linear and nonlinear variant, has thus eluded observation so far. 

The study of the propagation of classical and quantum pseudo-relativistic fields 
on {\em effective} curved spacetime dubbed  {\it analogue gravity} \cite{BLV} enables the simulation of many effects else inaccessible in the lab \cite{volovik2009universe}. 
After Unruh's seminal idea \cite{unruh},
various aspects of analogue gravity have been treated in particular within the realm of fluid dynamics. 
These comprise, {\it  inter alia}, analogues 
of black holes via gravity waves \cite{RalfBill,Weinfurtner,Euve,EuveII} and  in fluids of light \cite{Marino,Nguyen}, black hole radiation \cite{PhysRevLett.85.4643,Carusotto_2008,Macher,Gerace,Steinhauer16,Munoz}, inflation \cite{Schuetzhold,PhysRevLett.118.130404,Eckel}, the production of cosmological quasiparticles \cite{BLV2003PRA,CPP} and their degree of entanglement \cite{Busch,Robertson,Tian}, the Unruh and Gibbons-Hawking effects \cite{Fedichev,Reznik,Celi,Gooding}, black hole lasers \cite{Ted,Finazzi} 
superradiance \cite{Basak,Torres,Prain}, quasinormal black hole 
modes \cite{Berti,Richartz}, and GWs \cite{hartley2018analogue,PhysRevD.98.064049}.  
 

The Unruh paradigm of analogue gravity operates on the premise that a 
linearized theory on top of some background field \cite{BLV_normal} describes accurately the propagation of the fields on a spacetime background which is generally curved \cite{FischerVisser}.   
In what follows, we establish an acoustic analogue of GW memory 
within a {\em nonlinear} generalization of analogue gravity. 
Our approach is based on the Riemann equation of nonlinear acoustics,
and does not assume that perturbations can be linearized. 
We show that the acoustic analogue  of a generally nonlinear GW memory 
can be observed in a Bose-Einstein condensate (BEC),  for which  
the  nonlinearity of the underlying ``ether" 
is provided by the fluid-dynamical equations, and not by the Einstein equations. 
While GW memory analogues have been introduced before, within electrodynamics \cite{Bieri} and Yang-Mills theory \cite{Pate}, respectively, here we consider a system which both represents an inherently nonlinear theory, and is readily accessible in experiment.  

\section{Fluid-dynamical setup}
\subsection{Bose-Einstein condensates}
The dynamics of BECs is on the mean-field level 
captured by the Gross-Pitaevski\v\i~equation (GPE)  ($\hbar\coloneqq 1$, atomic mass 
$m\coloneqq 1$) \cite{pita2003bose}
\begin{equation}\label{gpe}
i\frac{\partial\psi}{\partial t}=\left(- \frac{\nabla^2}2 +V_{\rm trap} +g|\psi|^2\right)\psi,
\end{equation}
where {$\psi$ is the condensate wave function, 
$t$ is time, $V_{\rm trap}$ is the externally imposed trapping potential,} 
and $g$ is the two-body contact interaction coupling. 
To leave the salient physics unobscured,  we place the BEC in a box trap 
\cite{Gaunt}, and omit 
mention of 
$V_{\rm trap}$ in what follows.
We 
work in the Thomas-Fermi (TF) limit, wherein density variation 
length scales 
are much larger than healing length $\xi_c=(g \rho_0)^{-1/2}$, where
$\rho_0=\frac{N}{V}$,  
with $V$ 
volume and $N=\int d^3{\bm r} |\psi({\bm r},t)|^2 $ particle number. 
We use the Madelung transformation 
$\psi=\sqrt{\rho({\bm r}, t)}e^{i\Phi ({\bm r}, t)}$; {the density $\rho>0$, 
and the velocity potential $\Phi ({\bm r}, t)$ is a real function. 
It is easily shown that then the GPE is equivalent to the irrotational 
flow (excluding the singular quantized vortex lines) of a perfect fluid characterized by density and velocity
potential as defined by the Madelung transformation  
\cite{pita2003bose}.} 
Continuity equation, Euler equation, irrotationality condition and 
polytropic equation of state are, respectively, using  ${\bm v}=\nabla \Phi$,  
\begin{equation}
\begin{aligned}
\partial_t\rho+\nabla\cdot (\rho{\bm v})=0, &\qquad&  
\partial_t {\bm v}+{\bm v}\cdot \nabla{\bm v}=-{\nabla p}/{\rho}, 
\\
\nabla\times {\bm v}=0, &\qquad&   p=\frac{1}{2}g\rho^2. 
\end{aligned} 
\label{fluids}
\end{equation}
We then introduce 
perturbations 
on top of the constant density 
background at rest:
\begin{equation}
\begin{aligned} 
\rho({\bm r},t)=\rho_0+\delta\rho({\bm r},t), \qquad {\bm v}({\bm r},t)=0+{\bm v}({\bm r},t)=\nabla\delta\Phi,\\
c_{s}^2=\frac{\partial p}{\partial \rho}=g\rho_0+g\delta\rho({\bm r},t)\coloneqq c_{s0}^2+\delta c_{s}^2.
\end{aligned}
\label{sounds}
\end{equation}
{To set the stage for the following analysis of nonlinear acoustics in terms of the Riemann wave equation \eqref{rw}, 
we restrict ourselves to a strongly laterally confined quasi-one-dimensional (quasi-1D) BEC \cite{Goerlitz}. To remain
sufficiently general, we are now considering a polytropic equation of state, $p(\rho)=K\rho^\gamma$ (with $K>0$, and $\gamma>1$, where the BEC equation of state in \eqref{sounds} above has $\gamma=2$.
The Euler equation then, after one spatial integration, is taking the form
\begin{equation}\label{EI}
\partial_t\delta{\Phi}+\frac 12(\partial_x\delta\Phi)^2+\frac{c_s^2}{(\gamma -1)}-\frac{c_{s0}^2}{(\gamma -1)}=0.
\end{equation}
A partial time derivative taken of Eq.~\eqref{EI}, and a partial derivative with respect to $x$ gives the two equations 
\begin{eqnarray}
& \partial_t^2\delta\Phi+(\partial_x\delta\Phi)(\partial_x\partial_t\delta\Phi)+\frac{c_s^2}{\rho}\partial_t\delta\rho=0,\label{phitt}\\
& \partial_t\partial_x\delta\Phi+(\partial_x\delta\Phi)(\partial_x^2\delta\Phi)+\frac{c_s^2}{\rho}\partial_x\delta\rho=0.\label{phitx}
\end{eqnarray}
{{Substituting $v=\partial_x\delta\Phi$ in the continuity equation, we find 
\begin{equation}\label{tro}
\frac{1}{\rho}\partial_t\delta\rho=-\partial_x^2\delta\Phi-\frac{1}{\rho}\partial_x\delta\rho\partial_x\delta\Phi.
\end{equation}
Now, we substitute the above expression for $\frac{1}{\rho}\partial_t\delta\rho$ , 
replacing therein $\partial_x\delta\rho$ using Eq.~\eqref{phitx},  
into Eq. \eqref{phitt},  Thereafter, we use the relation \eqref{EI} to insert $c_{s}^2=c_{s0}^2-(\gamma -1)\left(\partial_t\delta\Phi+\frac{1}{2}(\partial_x\delta\Phi)^2\right)$.
}
{{As a result, we obtain a second order nonlinear partial differential equation in terms of only one variable, $\delta\Phi$:}
\begin{multline}
(-\partial_t^2+c_{s0}^2\partial_x^2)\delta\Phi=2(\partial_x\partial_t\delta\Phi )\partial_x\delta\Phi \\+(\gamma-1)(\partial_x^2\delta\Phi)\partial_t\delta\Phi 
+\frac{(\gamma+1)}{2}(\partial_x \delta\Phi)^2 \partial_x^2\delta\Phi .
\end{multline} 
}
{Thus, the nonlinearity of our problem is manifest in the full equation for phase fluctuation. 
Distinct from the case of linearized perturbations, $\delta\Phi$ does not satisfy a massless Klein-Gordon (KG) equation when nonlinearity is taken into account. In Ref.~\cite{datta2021analogue}, it has been established by us that such nonlinear perturbations back-react onto the background, and the definition of the {\em background} therefore changes. 
A new background solution is then defined (with an appropriate new notation for the perturbative expansion),  
by absorbing the nonlinear perturbation terms; $\rho\rightarrow \rho_{(0)}=\rho_0+\delta\rho$, ${\bm v}\rightarrow {\bm v}_{(0)}$, $\Phi\rightarrow \Phi_{(0)}=\Phi_{0}+\delta\Phi$. 
A higher frequency linear sound wave (the next order of perturbation in the velocity potential, $\Phi_{(1)}$) couples to the new background, and it then satisfies a massless Klein-Gordon (KG) equation, 
\begin{equation}\label{fwave}
\partial_\mu\left[ f^{\mu\nu}_{(0)}({\bm r},t)\partial_\nu\Phi_{(1)}({\bm r},t)\right]=0, 
\end{equation}
where we identified the effective spacetime contravariant tensor $f^{\mu\nu}_{(0)} 
\coloneqq\frac{\rho_{(0)}}{c_{s(0)}^2}\begin{bmatrix} -1 & -{\bm v}_{(0)}\\
-{\bm v}_{(0)} & c_{s(0)}^{2}\mathbb{1}-{\bm v}_{(0)}\otimes {\bm v}_{(0)} \end{bmatrix}$. } 

  
\subsection{Redefining the effective spacetime metric in 1+1D} 
{Evidently, the new background solution from nonlinearity results in a modification of the acoustic metric, 
$g_{\mu\nu} \rightarrow \mathfrak{g}_{\mu\nu}$ in $1+1$D.}
We use the conformal factor in the physical  $3+1$D spacetime,  
leading to the acoustic spacetime metric in 3+1D being given by 
$G_{\mu\nu}=\frac{\rho_{(0)}}{c_{s(0)}}\begin{bmatrix}
 -(c_{s(0)}^{2}-v_{(0)}^{2}) &-{\bm v}_{(0)} \\
 -{\bm v}_{(0)} & \mathbb{1}
 \end{bmatrix}$,  and then consider that transverse
directions are frozen out, see Appendix \ref{conform}. 
 Absorbing a constant conformal factor related to the uniform density of the background, we obtain the 
 metric we work with:    
 \begin{equation}\label{ngw}
 \mathfrak{g}_{\mu\nu}(x,t)=
 \begin{bmatrix}
-c_{s0}^2 & 0 \\
0 & 1
\end{bmatrix}
 +h_{\mu\nu}(x,t),\qquad \mu,\nu = t,x. 
 \end{equation}
The metric perturbations 
are  to second order given by 
\begin{eqnarray}
h_{tt}&=&-c_{s0}^2\!\!\left[\!\frac{(\gamma +1)}{2}\frac{\delta\rho}{\rho_0}\!+\!\frac{(\gamma^2-1)}{8}\!\left(\frac{\delta\rho}{\rho_0}\right)^2\right]\!+v^2,\label{htt}\\
h_{tx}&=&h_{xt}=-v\left(1+\frac{(3-\gamma)}{2}\frac{\delta\rho}{\rho_0}\right),\\
h_{xx}&=&\frac{(3-\gamma)}{2}\frac{\delta\rho}{\rho_0}-\frac{(3-\gamma)(\gamma -1)}{8}\left(\frac{\delta\rho}{\rho_0}\right)^2. \label{hxx}
 \end{eqnarray}
{Now, if the spatiotemporal variation of $\Phi_{(1)}$ is sufficiently fast, 
to be regarded as in the eikonal limit, then we are in the geometric regime \cite{BLV}, i.e., 
the massless scalar field $\Phi_{(1)}$ then follows a null geodesic given by 
\begin{equation}\label{dsgm}
ds^2=\mathfrak{g}_{\mu\nu}dx^\mu dx^\nu=0.
\end{equation}
 }
\subsection{Riemannian fluid dynamics} 
As established in the seminal work of Riemann on the exact 
description of shock waves \cite{Riemann1860},  
an analytical approach to nonlinear waves in perfect fluids is
furnished by Riemann invariants 
\cite{Nemirovski_Uspekhi,enflo2006theory}, associated to a 
Riemann wave equation \cite{sooderholm2001higher,hamilton1998nonlinear,kuznetsov1971equations}.
To render our argument as transparent as possible, we consider below {\it simple} waves, for which the constancy of one of the two Riemann invariants yields  (cf.~\S104 in \cite{Landau1987Fluid}) 
\begin{eqnarray}
\label{rhov}
\rho=\rho_0\left(1\pm(\beta-1)\frac{v}{c_{s0}}\right)^{\frac{1}{\beta-1}},
\end{eqnarray} 
where $\pm$ indicates propagation along the positive ($+$) or negative ($-$) $x$ direction and 
$\beta=\frac{\gamma+1}{2}$.  
Derived from the set of Eqs.~\eqref{fluids}  
 (cf.~\S101 in \cite{Landau1987Fluid}), the Riemann wave equation then reads 
\begin{equation}\label{rw}
\frac{\partial v}{\partial t}+(\pm c_{s0}+\beta v)\frac{\partial v}{\partial x}=0. 
\end{equation}
In a stable BEC with contact interactions, the constant 
$K=g/2>0$ and $\beta=\frac{3}{2}$.
The wave equation \eqref{rw}, a nonlinear first-order partial differential equation, 
specifies the solution of the nonlinear acoustics problem once an initial wave profile  $v=f(x)$ at $t=0$ is given.  
For small $v$ amplitudes, one has the linearized equation 
$\partial_t v +c_{s0}\partial_x v=0$ (assuming right-moving waves in what follows),  
which then leads to the conventional Unruh paradigm of analogue gravity.

We solve Eq.~\eqref{rw} by using the method of characteristics \cite{principles,enflo2006theory}, which employs constant $v$ curves in the evolving flow.  
For the characteristics 
curve specified by  $\frac{dv}{dt}=0$, 
we seek the contour map 
on the $x-t$ plane, each contour curve representing a fixed $v$,  
\begin{align}
\left.\frac{dx}{dt}\right|_v=c_{s0}+\beta v, \qquad 
x=\left(c_{s0}+\beta v\right)t+\xi, 
\label{characteristics}
\end{align}
and $ v=f(\xi)$ depends on the characteristics parameter $\xi$, $\ldots |_q$ denoting constant $q=\{ v, \delta\rho,\ldots\}$.  
We also have, from  \eqref{rhov} and \eqref{rw}, $\frac{dx}{dt}|_v=\frac{dx}{dt}|_{\delta\rho}= c_{s0} + \beta v = c_s+v$. 
{Since the Riemann wave solution fully incorporates nonlinearity, redefining the background here 
yields $\rho_0\rightarrow \rho_{(0)}$, $v\rightarrow v_{(0)}$. From now on, we consider the Riemann wave as providing the background, and switch to this notation. 
We also note here that  
Ref.~\cite{marino2016emergent} experimentally realizes such an 
emergence of an effective metric in the geometric limit, 
governed by a nonlinear simple wave solution of the Riemann wave equation in photon fluids.}  

\subsection{Avoiding shock waves}
Considering an initially ($t=0$) monochromatic wave profile, 
$v_{(0)}=A\cos k\xi$, 
we find that at $t=t_{\rm shock}=\frac{1}{\beta Ak}$, the derivatives of $v$ with respect to $x$ and $t$  diverge at zeros with negative slope of $v$,  
see Appendix \ref{shock}. 
For $t>t_{\rm shock}$ (the first instant at which the wave profile becomes infinitely steep),  
$v$ and (thus $\delta\rho$), becomes multivalued, i.e., at a given  instant of time, a fixed position can correspond to different values of $v$. 
The characteristic lines of Eq.~\eqref{characteristics} start to intersect with each other, 
resulting in shock waves ($\equiv$ discontinuities in $v$ and $\rho$}) \cite{enflo2006theory}, studied for BECs in \cite{kamchatnov2004dissipationless,meppelink2009observation}. 
We restrict ourselves to nonlinear acoustic flow without 
shock waves developing, i.e, consider times 
$t<t_{\rm shock}$. 
Imposing $\beta A<\frac{c_{s0}}{2\pi}$ implies 
$\frac{2\pi}{c_{s0}k}< t_{\rm shock}$ 
\cite{enflo2006theory,Landau1987Fluid}.  
For the acoustic GW memory analogue, we need a wave train
passing through the BEC cloud without encountering a shock wave discontinuity. 
Indeed, for times of order $t_{\rm shock}$ a perturbative theory breaks down. 
{When parts of the wave profile become steeper over time, evidently the TF approximation fails. 
However, the quantum pressure term in the GPE in fact prohibits such a discontinuity to emerge, 
and it has been shown that instead an oscillation pattern in density forms 
as $t$ approaches $t_{\rm shock}$ \cite{PhysRevA.69.043610}.}
\section{Acoustic GW memory}
Owing to general covariance, the movement of a single test particle caused by a GW can be nullified to first order in $h_{\mu\nu}^{TT}$ by a proper choice of coordinate system \cite{carroll} and permanent relative displacements of at least two test particles need to be measured to ascertain the impact of a GW, also see Appendix \ref{differences}. This change of distance 
occurs due to a permanent change in the TT component of $h_{\mu\nu}$. 
When the oscillatory part of a GW with memory vanishes as $t\rightarrow\infty$, i.e., a freely falling test mass does not feel a gravitational acceleration due to GW, then the relative distance between two test particles ultimately settles to a permanent change due to the presence of a remaining nonoscillatory part of the signal. 

{In our nonrelativistic laboratory setup}, the observer 
has the Newtonian notions of distance and absolute time.   
In the context of our acoustic GW analogue, we therefore have no obvious choice of geodesic test particles at our disposal to determine the impact of a GW.
We therefore use as our primary tool {\em fiducial} particles (FPs). 
\begin{figure}[t]
  \centering
   \vspace*{0.5em}
\includegraphics[scale=0.16]{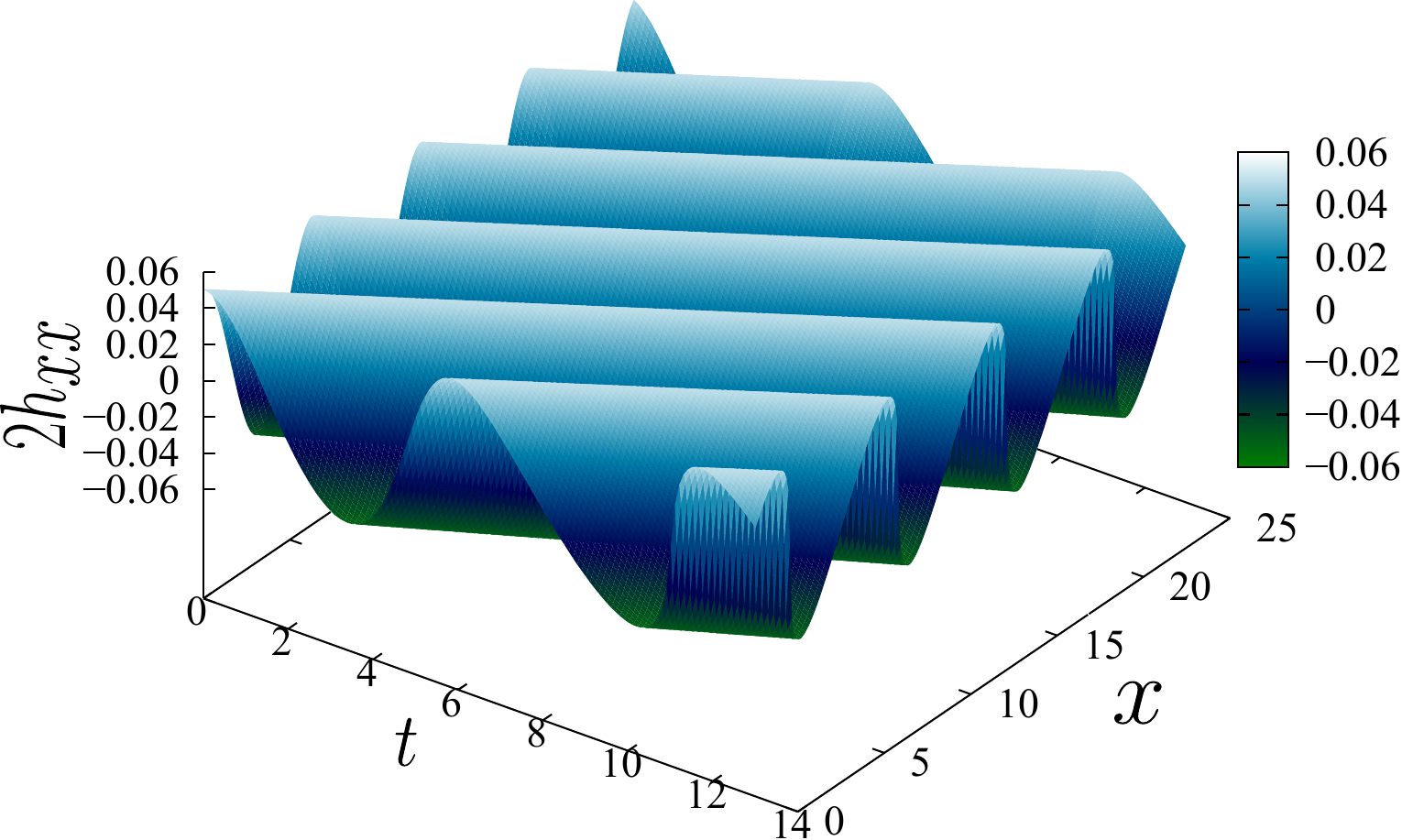}  
\includegraphics[scale=0.16]{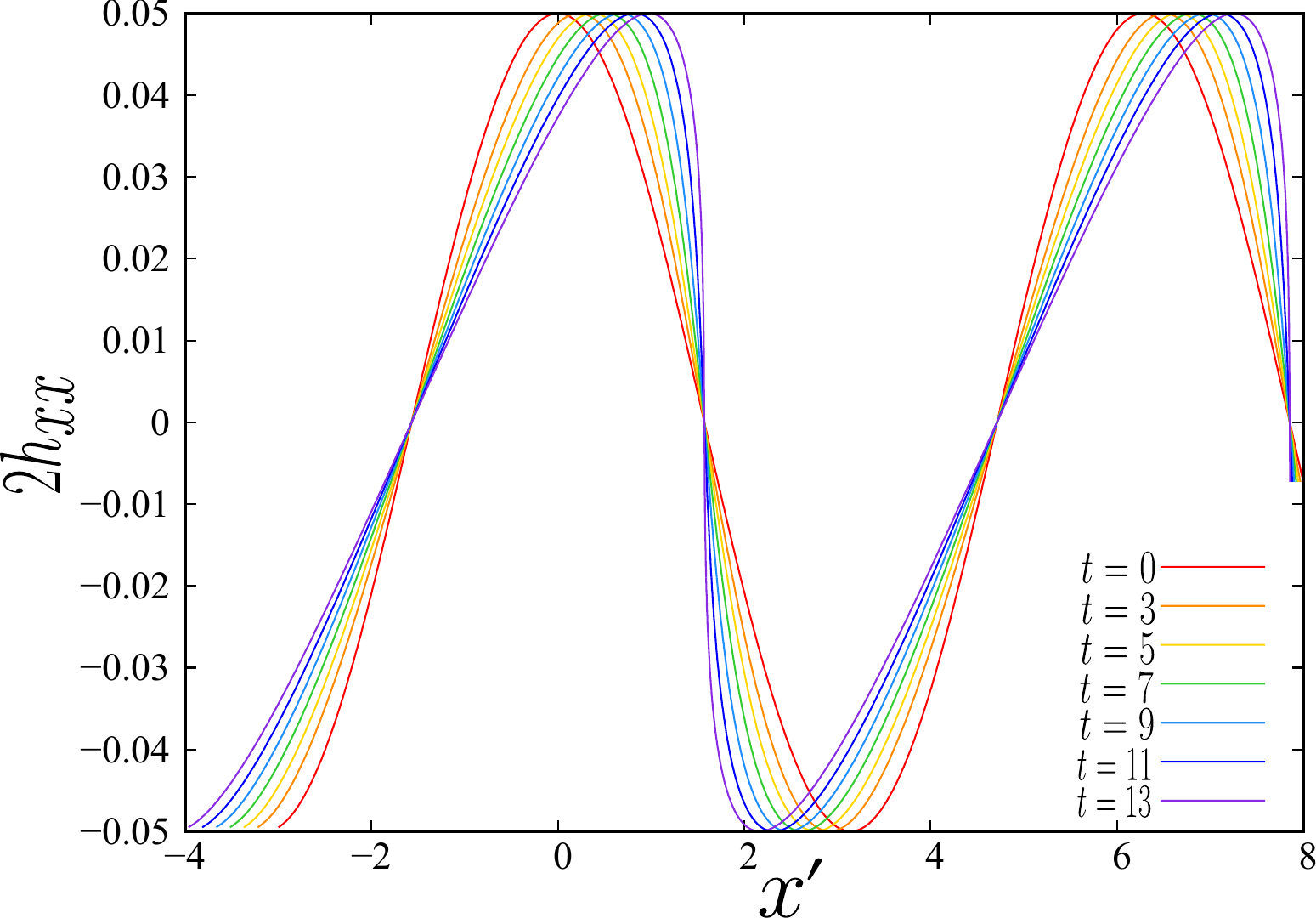}
   \caption{Schematic illustration of the GW amplitude  
   $h_{xx}~(=\frac{v_{(0)}}{2c_{s0}})$  for $\beta=\frac 32$ ($\gamma=2$). We choose $v_{(0)}=A \cos kx $;   
   $x,x'=x-c_{s0}t$  and $t$ are in units of  
  $1/k$ and $1/(c_{s0}k)$, respectively. 
   {Left}:    $h_{xx}$ in the $x,t$ plane ($A=0.05$). 
{Right}: Temporal change of initial cosine profile; greater value of $h_{xx}$ corresponds to smaller speed $v_{(0)}$, clearly visible in the 
frame  $(x',t)$ comoving at $c_{s0}$. Here, $t_{\rm shock}=\frac{40}{3}$.}
\label{fig1}
\vspace*{-0.5em}
\end{figure}
We define the FPs as particles which move with the flow. 
As the FP moves $\xi$ changes; a specific value of $\xi$ gives a characteristic curve, i.e., the sound wave trajectory. The equation of motion is just $\frac{dx}{dt}=v_{(0)}$, with the initial condition $x=\xi_i$ at $t=0$. 
We find $v_{(0)}=v_{(0)}(\xi)=v(x,t)$ by solving for $\xi=\xi(x,t)$ from Eq.~\eqref{characteristics}. 
For a sound (null) ray, $\frac{dx}{dt}=(\pm c_{s(0)}+v_{(0)})$. For $v_{(0)}>0$, $v_{(0)}<c_{s(0)}+v_{(0)}$, for $v_{(0)}<0$, $v_{(0)}>-c_{s(0)}+v_{(0)}$, 
we have already established in the above 
that $v_{(0)}<c_{s(0)}$ and $c_{s(0)}=c_{s0}+\beta v_{(0)}$.  
We conclude that $\forall (x,t)$ the FP lies inside an intersecting pair of null lines.
To visualize the change in shape of the wave profile, 
due to the nonlinearity, $\beta \neq 0$, in the Riemann Eq.~\eqref{rw},  
we construct a comoving reference $(x',t)$ frame moving at $c_{s0}$ along the positive $x$ axis,
cf.~the right plot in Fig.~\ref{fig1}.
To establish the analogy, we need, as explained in the above, to realize a permanent change in the metric  
after the GW has passed. For simplicity, we choose to work with $v_{(0)}=f(\xi)=B$ for $0<\xi<\frac{n\pi}{2k}$; $v_{(0)}=f(\xi)=A\cos k\xi+B$ for $\frac{n\pi}{2k}\leq\xi\leq\xi_i$ with $n$ being a positive odd integer, $f(\xi_i)=0$. This wave profile of $v_{(0)}$ causes a permanent 
 change in $h_{\mu\nu}$ once  the 
 the GW has passed. 
We note that even if one were to consider a FP following the geodesic equation of phonons and not moving with the flow, constant $h_{\mu\nu}$ would produce no relative acceleration between two FPs.  
The permanent metric changes in the metric components due to the presence of such a constant part at the trailing end of the GW signal then are, up to second order in the velocity perturbation constant $B$, 
 \begin{eqnarray}\label{memoryh}
\Delta h_{tt}&=&\left(-\frac{(\gamma +1)}{2}c_{s0}B+ \frac{3-\gamma}{4} B^2\right), \nn
\Delta h_{tx}&=& \left(-B-\frac{(3-\gamma)}{2}\frac{B^2}{c_{s0}}\right),\nn  
\Delta h_{xx}&=&\left(\frac{(3-\gamma)}{2}\frac{B}{c_{s0}}+\frac{(3-\gamma)(2-\gamma)}{4}\frac{B^2}{c_{s0}^2}\right) . 
\end{eqnarray}
Distinct from the real transverse GW, we have a longitudinal wave in the BEC medium as the acoustic analogue of a GW 
in our fluid-dynamical set up. Here the component $h_{xx}$ plays a similar role to that of the transverse component $h_+$ in a real GW. 
We observe here that the nonlinear part of analogue GW memory, in distinction to its
Einstein-gravitational counterpart, is also nonvanishing in the (however hypothetical, because 
thermodynamically unstable) case when $\beta=0$ ($\gamma=-1$). 
This is due to the relation~\eqref{rhov} ($v_{(0)}^2$ terms do not vanish for $\beta=0$) and thus reflects the 
underlying structure of the medium (a polytropic gas), in which the analogue GW propagates. 
We also note that $\beta =2$ ($\gamma=3$) is a special case. Using Eq.~(15), which for $\beta=2$ yields a {\em linear} relation between density and velocity perturbations, one readily shows that $\Delta h_{xx}=0$ to any order in $B$, and that in $\Delta h_{tt}$ and $\Delta h_{tx}$ all terms higher than linear in $B$ exactly vanish.


\section{Experimental setup}
\subsection{Phase-imprinting memory analogues}
To provide a memory-related quantity which can be experimentally accessed by a quantum-optical  observer in the laboratory,     
we consider specifically a homogeneous cloud of ultracold 
atoms in a cylindrical box trap \cite{Gaunt} 
with radius $R\,(\ll \xi_c)$ and length $L\,(\gg \xi_c)$, 
see~Fig.~\ref{figNonlinearmain}.  
We suggest to employ the phase imprinting technique, readily  
available in the quantum optical setup of ultracold gases
\cite{Denschlag,Leanhardt}, to implement a specific GW profile corresponding to the initial velocity perturbation. 
We create a spatial variation in the phase of the initial 
condensate wave function, within the source region ($S$) of length $l$, by red-detuned 
laser light turned on for a short duration $T$ (as short as to stay within the Raman-Nath regime of simple diffraction). 
The superfluid phase pattern $\Phi(x) \propto I(x)/\delta$, corresponding to the velocity pattern discussed above, 
where
$I(x)$ is laser intensity and $\delta$ detuning from resonance, is then imprinted in $S$,  where the phase profile is chosen as 
\begin{eqnarray}\label{PhiProfile}
\Phi(x) &= &Bx+C,\quad{\mbox{for}~0<x<\frac{n\pi}{2k}},\nonumber\\
\Phi(x) &=& \frac Ak\sin(kx)+Bx+C, {\quad\mbox{for}~\frac{n\pi}{2k}\leq x<l}, 
\end{eqnarray} 
{{where $n$ is an odd integer.} 
Here, we put $A,B,C\geq 0$, and the constant $C$ is chosen such that $\Phi (x)>0$ within 
S (red-detuning $\forall\, x\in S$), {cf.~\ref{fig2} for an illustration of the resulting $h_{xx}$ component. 
{{We note that in the $h_{xx}$ series, using in \eqref{memoryh} the relation \eqref{rhov} (with a $+$ sign), 
the second power term in $v$ does not appear for a  BEC which has $\gamma=2$, 
and  we have $h_{xx}={v}/{2c_{s0}}$ up to  cubic order.}}

\begin{figure}[t]
  \centering
   \vspace*{0.5em}  
\includegraphics[scale=0.3]{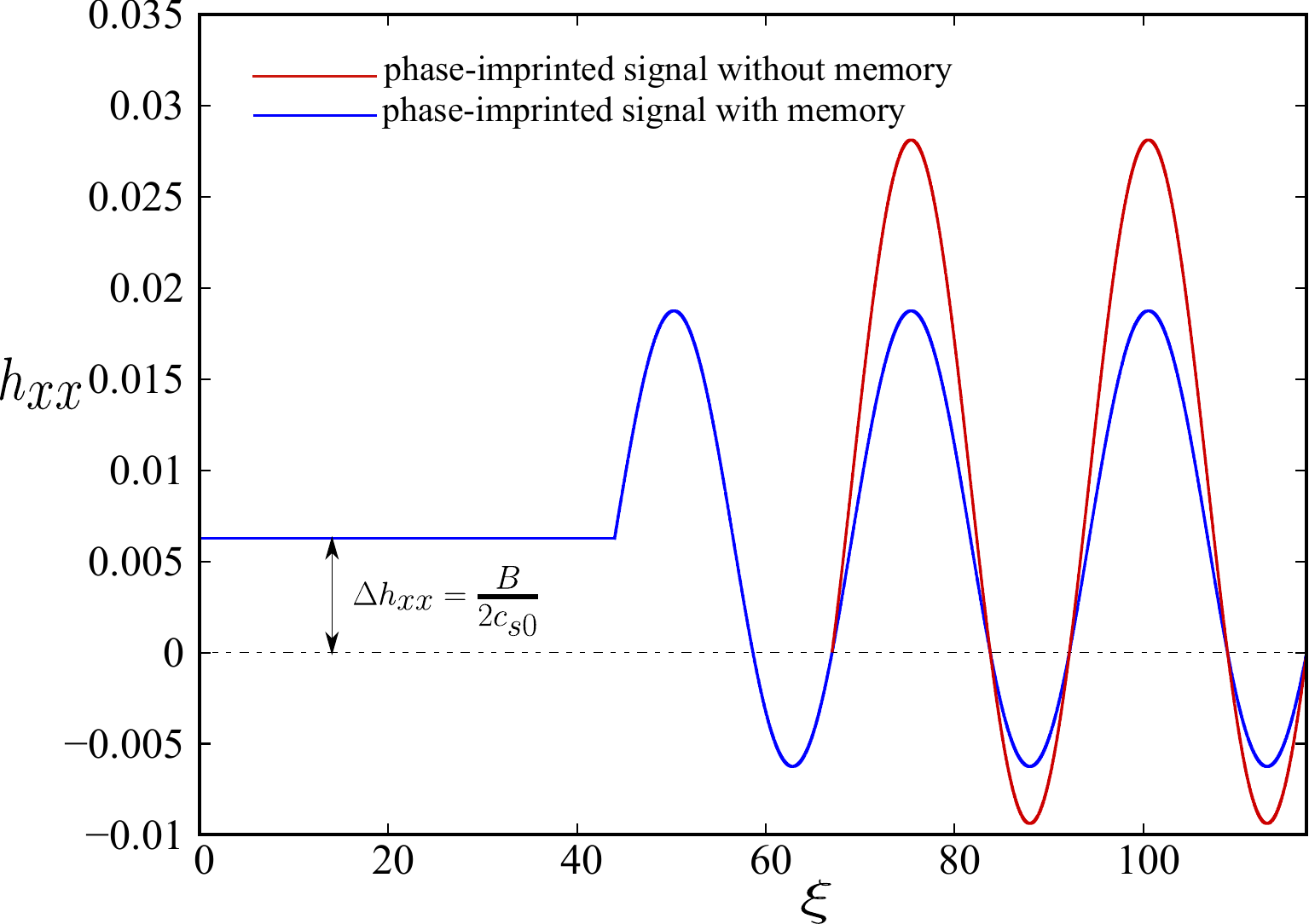}
   \caption{The initial $t=0$ metric $h_{xx}$ {according to Eq.~\eqref{hxx}}
   in the source region, reasulting in a GW with memory
   [blue curve, for $B=A/2$, $A=0.025$, and $n=7$ in Eq.~\eqref{PhiProfile}], and without memory  
  [red curve, for which we set $\Phi(x)=0.0375\cos(0.25x)+0.01875x$ for $\frac{64\pi}{3}\leq x\leq \frac{112\pi}{3}$; and $\Phi(x)={\rm constant}$ for $0\leq x<\frac{64\pi}{3}$]. We put $k=0.25\xi_c^{-1}$; $B/A=1/2$, $A=0.025$ in units of $c_{s0}$, and $\xi$ is in units of the healing length $\xi_c$. The length of the source region is assumed to be $l=\frac{112\pi}{3}\xi_c$. Here,  $\xi_c=1.8\,\mu$m (as in Ref.~\cite{Munoz}). 
For the profile with memory, this results in $l=0.21$\,mm, less than the maximum allowed value of $l=0.23$\,mm, which would lead to shock formation. 
}
\label{fig2}
\vspace*{-0.5em}
\end{figure}  

To simplify the calculation, we choose to work with the above nonsmooth ansatz for $\Phi$ (noting that experimentally any required smooth profile  can be engineered). The thus created  bipartite 1D configuration, cf.~Fig.~\ref{figNonlinearmain},   produces simple waves \cite{Landau1987Fluid}. {At $t=0$, $v_{(0)}(\xi_i)=v_{(0)}(l)=0\Rightarrow B<A$; this is how the constant magnitude of the profile at its edge is bounded by the amplitude in the oscillatory part in the front.} 
{Considering $^{87}\!$Rb as a 
BEC  as employed in the Hawking radiation experiment of Ref.~\cite{Munoz}, 
with healing length $\xi_c=1.8\,\mu$m (geometrically averaged 
between upstream and downstream regions relative to the horizon), 
we also have $c_{s0}=0.39$\,mm/sec. 
We consider that the amplitude $A/c_{s0}=0.025$, and that the wavevector $k=0.25\xi_c^{-1}$ is of order the 
dominant Hawking mode in Ref.~\cite{Munoz}. 
Such a value of $k$ can be achieved by phase-imprinting with an infrared laser. }

{{We now aim at generating the memory associated to $h_{xx}$, similar to what we expect to have in the $h_+$ signal
for a real GW in relative proportion to the highest magnitude of the oscillatory part of the GW.}
The maximum value of $h_{xx}$ in the oscillatory part of the phase-imprinted simple wave in 
Eq.~\eqref{PhiProfile} 
is $\frac{(A+B)c_{s0}}{2}$, and 
the expression \eqref{memoryh} yields $Bc_{s0}/2$ (for a BEC $\gamma =2$).
Therefore $\frac{B}{A}=\frac{1}{n'}$ with $n'\geq1$, produces a signal having a memory part $1/(1+n')$th of the maximum value in the oscillatory part. We denote this ratio $w=1/(1+n')$. Hence in the specific phase-imprinting example 
\eqref{PhiProfile}, $w\leq1/2,~\because n'\geq1$. For example, referring here to Fig.~1 in the review of Ref.~\cite{favata2010gravitational}, which displays a typical gravitational wave signal having memory,  
$w$ is there roughly $1/3$, and thus the corresponding value of $n'$ used for phase-imprinting is about two.



In the region of observation (RO), the observers in the lab take density images by absorption imaging at discrete times. A constant density tail of the wave in the RO, as depicted in the Fig.~\ref{figNonlinearmain} represents the memory of the signal. Eq. \eqref{rhov} gives this constant density shift, which is, up to order $B^2$,  
\begin{equation} 
\Delta\rho=\rho_0\left(\frac{B}{c_{s0}}+\frac{(3-\gamma)}{4}\frac{B^2}{c_{s0}^2}\right)~\mbox{with }B<A<c_{s0}.
\label{deltarho}
\end{equation}
An auxiliary condition is that discontinuities due to a shock developing have to be avoided. 
The mean speed of the signal equals the speed of the constant part of the wave train, i.e., $c_{s0}+\beta B$, as follows from Eq. \eqref{characteristics}.   
The wave train's oscillatory part encounters a shock at $t=t_{\rm shock}=\frac{1}{\beta A k}$ (cf.~Appendix \ref{shock}). To avoid discontinuities, we thus design the experiment such that $\frac{l}{(c_{s0}+\beta B)}<t_{\rm shock}$.  
{With the aforementioned parameters from \cite{Munoz}, $t_{\rm shock}=0.57$\,sec; for $B/A=1/2$, $A=0.025$, $l<0.23$\,mm. Absorption images are then taken at times $t<t_{\rm shock}$. }

\begin{figure}[t]
  \centering
\includegraphics[scale=0.16]{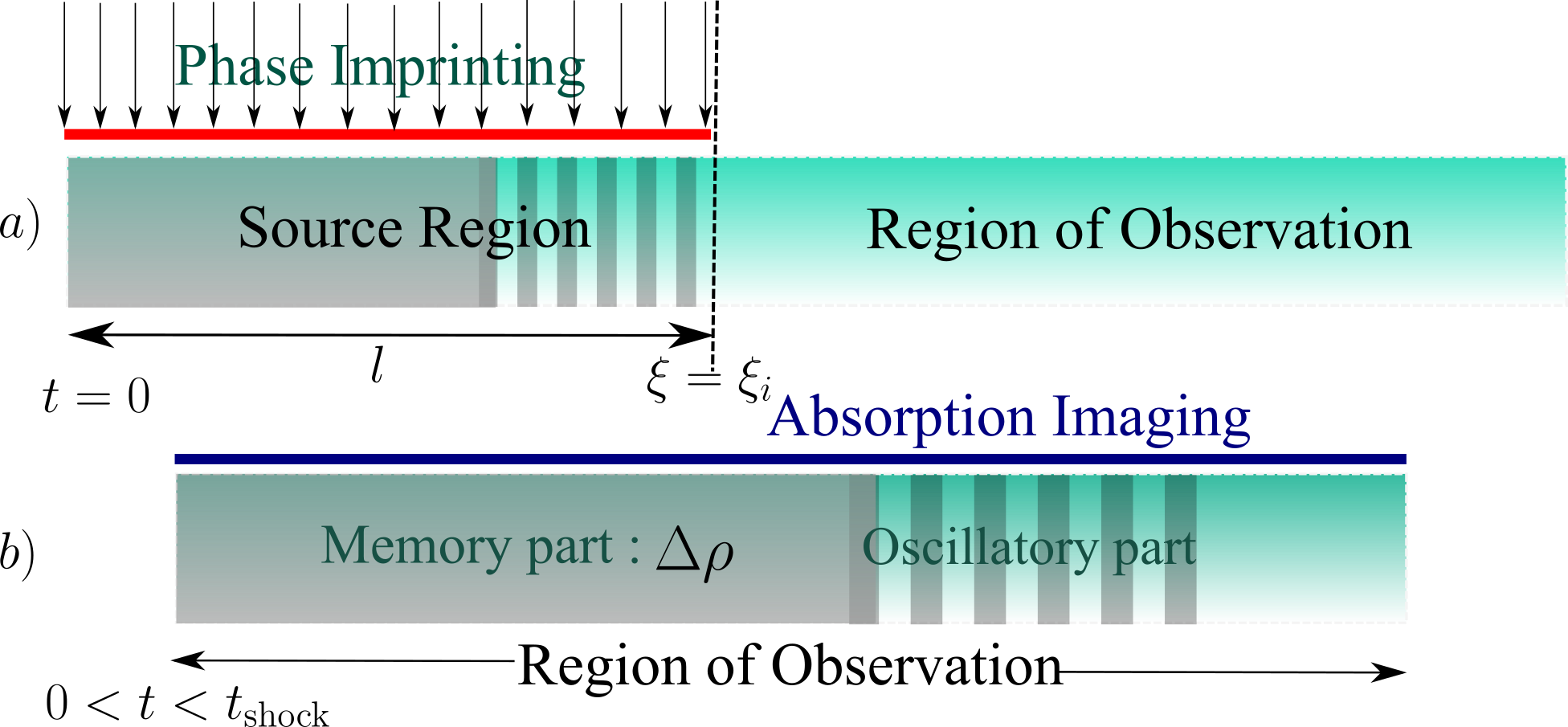}
    \caption{{\em Observing the acoustic analogue of GW memory in a quasi-1D BEC.} The grey shading represents the schematic density variation.
   a) The GW signal with memory is produced in the {\it source region} by phase imprinting at $t=0$, b) {\it Zoomed-in region of observation} at a time of observation less than $t_{\rm shock}$. The uniform grey shading represents constant density [increase or decrease depending on the sign of $B$ in Eq.\eqref{deltarho}]. The memory of the signal is captured by absorption imaging  
   when the oscillatory part is already completely residing within the region of observation.
    \label{figNonlinearmain}}
    \vspace*{-0.5em}
\end{figure}

\subsection{A proposal for simulating more realistic signals} 
{In the case of a real GW, the transverse part of $h_{\mu\nu}$, $h_+$, is an observable
and the memory part of it, e.g., depends on the total mass of the binary system, its reduced mass ratio, the distance between source and observer, the angle between source and direction of observation, and on the time evolution of the dominant source quadrupole moment from $t\rightarrow -\infty$ to the retarded time of detection, see for a discussion Ref. \cite{favata2009s}.
The latter detection time can in principle be arbitrarily large, distinct from our analogue nonlinear acoustics model. 
In our setup, we
have an {\it indirect} memory signal because we are not measuring permanent changes in the acoustic metric (however see below).
This indirect observable is a change in density, and the signal is longitudinal. We can however in principle engineer a more realistic profile (possible by phase imprinting) 
having an inspiral part and a ringdown part along with a memory part (referring here again to Fig.~1 of Ref. \cite{favata2010gravitational}) by knowing the Fourier spectrum of the signal, where an obvious rescaling of the length scales and time scales of the real GW signal will need to be performed.} 

\subsection{Direct method to access metric changes}
{{As stated in the above, measuring $\Delta\rho$ is an indirect method of observing the 
permanent change in the metric as expressed of Eq. 18. 
We now  propose to access more directly the permanent change in the 
metric by observing the time it takes for sound rays to propagate in the presence of the analogue GW signal.}
Let us assume a sound ray propagating on top of the simple Riemann wave in the region of observation. 
From Eq.~\eqref{dsgm}, the sound cone will have different opening angles in the presence of the GW
which may be represented as 
\begin{equation}
\frac{c_{s0}dt}{dx}=\frac{-\frac{h_{tx}}{c_{s0}}\pm \sqrt{\frac{h_{tx}^2}{c_{s0}^2}-(1+h_{xx})(-1+\frac{h_{tt}}{c_{s0}^2})}}{(-1+\frac{h_{tt}}{c_{s0}^2})}.
\end{equation}
Putting $h_{\mu\nu}$ to zero, one verifies that $\frac{c_{s0}dt}{dx}=\pm 1$. As the sound ray passes, in the memory region we have $h_{\mu\nu}=\Delta h_{\mu\nu}$. With the values chosen in the above, $B/A=1/2$, $A=0.025$,
we obtain for the sound ray going in the direction of GW propagation, $\frac{c_{s0}dt}{dx}\simeq 0.98$, and for the sound ray going opposite to the direction of  the GW, $\frac{c_{s0}dt}{dx}\simeq -1.01$.
Therefore, detecting the travel times of sound pulses in the eikonal approximation 
in the presence of the GW memory potentially yields direct access to the permanent change in the metric incurred by GWs with memory. 

\section{Conclusion}

Extending the linearized analogue gravity paradigm to nonlinear acoustics,  
we have demonstrated that an acoustic analogue of 
GW memory can be simulated in a BEC. 

For clarity, we delineate and reiterate now again the primary 
differences to Einsteinian GW memory, also see Appendix \ref{differences}:  
(a) Our GW analogue is a longitudinal (instead of transverse) wave and lives in 1+1D;   
(b) 
it is imposed locally by a quantum optical experiment, 
whereas an Einsteinian GW in 3+1D 
contains the physical distance $r_s$ from source to detector; 
(c) The Einsteinian GW 
and its memory is detected by relative geodesic acceleration. 


The acoustic GW memory analogue, 
on the other hand,   
offers an alternative perspective on GW memory 
allowing to disentangle its fundamental aspects.
For GW memory and its acoustic analogue, e.g., the impact of the effective 
equation of state of the underlying ``ether"  on the memory signal 
can be studied. 
The acoustic memory analogue can potentially play a role akin  
to Hawking radiation analogues, which led to the clear
recognition that distinct from black hole entropy, radiation 
is of a kinematical rather than Einstein-dynamical origin \cite{Visser}. 
{Checking the ``ether" dependence can for example be performed by phase-imprinting 
in ultracold gases waveforms akin to those of real gravitational waves, and then checking 
whether the memory signal response of real and acoustic GW memory 
displays a qualitatively similar behavior.}
Finally, ultracold gases 
lend themselves naturally to 
the simulation of the impact of   
exotic solitonic or vortical structures \cite{Denschlag,Leanhardt} in the GW train 
onto characteristic features of the memory signal. 

Looking further ahead, we expect a significant number of  
novel insights gained from the inherently nonlinear dynamics of fluids for simulating curved spacetimes. 
Apart from the concrete example of BECs discussed in the above, this applies equally well, e.g., to shallow water or fluids of light. 

\section{Acknowledgments} 
This work has been supported by the National Research Foundation of Korea under 
Grants No.~2017R1A2A2A05001422 and No.~2020R1A2C2008103.
It is dedicated to the memory of Renaud Parentani, and his many important and thorough contributions to analyzing classical and quantum fields in the effective curved spacetimes of fluids.

\begin{appendix}

\section{The quasi-1D gas and general dimensionality}\label{conform}
In a quasi-1D system along the $x$ axis, the dynamics along $y$, and $z$ axes is by definition frozen out, $v^{1}=v_{(0)}$, $v^{2}=v^{3}=0$. 
We may then write 
\begin{equation}
G_{\mu\nu}({\bm{r}},t)\equiv\frac{\rho_{(0)}}{c_{s(0)}}\begin{bmatrix}
 -(c_{s(0)}^{2}-v_{(0)}^{2}) & -v_{(0)} & 0 & 0\\
-v_{(0)} & 1 & 0 & 0\\
0 & 0 &  1 & 0\\
0 & 0 & 0 & 1
\end{bmatrix}
\end{equation}
In quasi-1D, the variables $\rho_{(0)}$ (hence $c_{s(0)}$), $v_{(0)}$ are functions of $(x,t)$. Therefore, the transverse dimensions $y$, $z$ does not play any role in the fluid dynamics. 
Therefore, we can construct an effective metric $\mathfrak{g}^c_{\mu\nu}(x,t)$:
\begin{equation}
\mathfrak{g}^c_{\mu\nu}(x,t)=\frac{\rho_{(0)}}{c_{s(0)}}\begin{bmatrix}
 -(c_{s(0)}^{2}-v_{(0)}^{2}) &-{ v_{(0)}} \\
 -{ v_{(0)}} & 1
 \end{bmatrix}, 
\end{equation}
where $\mathfrak{g}^c_{xx}=G_{yy}=G_{zz}$. In quasi-1D, one can indeed work with $G_{\mu\nu}$, we however 
choose to work with $\mathfrak{g}^c_{\mu\nu}(x,t)$ for convenience.  

In the main text, we consider this quasi-1D set up, embedded in 3D spatial dimensions. The acoustic metric gets reduced to $1+1$D 
from $3+1$D due to the circular (trapping) symmetry. In this section, we show that the whole analysis can be performed starting from $d+1$D embedding dimensions for general positive integer $d\geq 2$. We work with the conformal factor in $d+1$D \cite{BLV}. Indicating the metric with inclusion of the conformal factor by the superscript $c$, and where we assume already 
that transverse dimensions are frozen out, we have 
\begin{eqnarray}\label{ngwd}
& \mathfrak{g}^c_{\mu\nu}(x,t)=\left(\frac{\rho_{(0)}}{c_{s(0)}}\right)^{\frac{2}{d-1}}\begin{bmatrix}
 -(c_{s(0)}^{2}-v_{(0)}^{2}) &-v_{(0)} \\
 -v_{(0)} & 1
 \end{bmatrix}.
 \end{eqnarray}
The conformal factor, $\left(\frac{\rho_{(0)}}{c_{s(0)}}\right)^{\frac{2}{d-1}}=\left(\frac{\rho_0}{c_{s0}}\right)^{\frac{2}{d-1}}\left(1+\frac{\delta\rho}{\rho_0}\right)^{\frac{3-\gamma}{d-1}}$ by using a barotropic equation of state. The nonzero constant $\left(\frac{\rho_0}{c_{s0}}\right)^{\frac{2}{d-1}}$ factor can be absorbed into a redefined metric.  
We find
 \begin{equation}
 \mathfrak{g}_{\mu\nu}(x,t)=g_{\mu\nu}+h_{\mu\nu}(x,t), 
 \end{equation}
 where  the background metric is
 \begin{eqnarray}
 & g_{\mu\nu}\coloneqq\begin{bmatrix}
-c_{s0}^2 & 0 \\
0 & 1
\end{bmatrix},
 \end{eqnarray}
 and 
the perturbation metric $h_{\mu\nu}(x,t)$ is, considering~terms~up~to~second~order, 
\begin{widetext}
\begin{eqnarray}
& h_{tt}=-c_{s0}^2\left\{\left(\frac{\gamma(d-2)-d+4}{d-1}\right)\frac{\delta\rho}{\rho_0}+\left(\frac{(\gamma(d-2)-d+4)(\gamma(d-2)-2d+5)}{2(d-1)^2}\right)\left(\frac{\delta\rho}{\rho_0}\right)^2\right\}+v^2,\label{httd}\\
&h_{tx}=h_{xt}=-v\left(1+\frac{(3-\gamma)}{(d-1)}\frac{\delta\rho}{\rho_0}\right),\qquad 
h_{xx}=\left(\frac{3-\gamma}{d-1}\right)\frac{\delta\rho}{\rho_0}+\frac{(3-\gamma)(4-\gamma -d)}{2(d-1)^2}\left(\frac{\delta\rho}{\rho_0}\right)^2. \label{hxxd}
 \end{eqnarray}
 \end{widetext} 
{Note that the perturbed fluid velocities $v$ and $v_{(0)}$ are identical because we start with a static background.}
\section{Shock waves}\label{shock}
We consider for simplicity an initially monochromatic wave travelling along the $x$ axis, which has $v_{(0)}=A\cos k\xi$. 
Hence, we find from the Eq.~\eqref{characteristics} in the main text 
\begin{equation}\label{nr}
x=\left(c_{s0}+\beta A\cos k\xi \right)t+\xi.
\end{equation}
Therefore, we also find, by using \eqref{nr}, 
\begin{eqnarray}
& \partial_t v_{(0)}=-Ak\sin (k\xi)\partial_t \xi=-A k\frac{(c_{s0}+A\cos k\xi)}{\left(\beta Akt\sin k\xi-1\right)}\sin k\xi,\nonumber\\
& \partial_x v_{(0)} =-Ak\sin (k\xi)\partial_x\xi=\frac{Ak\sin k\xi}{\beta Akt\sin k\xi-1}. \label{vxrw}
\end{eqnarray}
Using the above two equations, one can check that the Riemann equation \eqref{rw} is satisfied. One can also see that at $t=t_{\rm shock}$, $\partial_t v_{(0)}$ and $\partial_x v_{(0)}$, become infinity. Different $\xi$ values correspond to different $t_{\rm shock}$, and the minimum value of $t_{\rm shock}$ is given by 
\begin{equation} \label{tinfy}
t_{\rm shock}=\frac{1}{\beta Ak},
\end{equation}
which is inversely proportional to the amplitude of velocity perturbation, $A$. 
Evidently, the expression Eq.~\eqref{tinfy}, remains unchanged if we add a constatnt to $v_{(0)}$, i.e., $v_{(0)}=A\cos k\xi +B$. For $t>t_{\rm shock}$, the density and velocity of the medium becomes multivalued, i.e., at a single coordinate $(x,t)$, more than one solution in $v_{(0)}$ and $\rho_{(0)}$ exists. This singular situation for $t>t_{\rm shock}$ is handled by introducing discontinuities in density and velocity, as discussed in \cite{Landau1987Fluid,enflo2006theory}. Discontinuous density and velocity solutions are  generally called shock solutions.

\section{GW memory in general relativity vs analogue memory} \label{differences} 

\subsection{Memory effect in general relativity}
Owing to general covariance, the movement of a single test particle by GW can be nullified 
to first order in the (traceless-transverse gauge) perturbation 
$h_{\mu\nu}^{TT}$ by a proper choice of coordinates \cite{carroll}, and relative displacements of at least two test particles have to be employed to ascertain the impact of a GW \cite{weinberg1972gravitation,carroll}. 
To measure the motion of a {\em single} test particle, we would then need to compare its displacement to neighboring objects (in some definite reference frame) which are also moving with the GW. 


The memory effect is characterized by the permanent change in separation between two freely falling particles (ideal detector) after a GW train has passed. The vectorial separation between two freely falling particles initially at rest relative two each other separated by a distance $l^k$, changes due to the GW field. The vectorial change $\delta l_j$ is given by (as derived from the equation of geodesic deviation) \cite{carroll}
\begin{equation}
\delta l_j=\frac{1}{2}h_{jk}^{TT} l^k.
\end{equation}
After a GW causing no memory has passed, $\delta l_j$ becomes zero, and the test particles come back to their original positions, which is not the case for a GW presenting memory. 
We note in this regard that the precise nature of GW memory, as originally proposed in 
\cite{Zeldovich}, has been debated.
Arguments have been put forward that after the (plane) GW wave train has passed,  
two given freely falling test particles do not come to rest again, as stated by \cite{Zeldovich}, 
but retain a finite velocity relative to each other \cite{Grishchuk_1989,Bondi,Zhang}.

\subsection{Memory effect in the analogue model}
In analogue models, general (coordinate) 
covariance is absent.  In other words, for our analogue, the observers in the lab are not affected by the 
 GW which is a sound wave propagating in the BEC medium. 
In the analogue models based on the description of fluid motion in Newtonian framework, 
the notion of time and space is absolute \cite{newton1687},  which is simply reflected in the preferred
frame of the (Newtonian) lab in which the experimentalist carries out his or her work. 

To build our analogy, we defined in the main text fiducial test particles which move with the flow. 
 We note here that one can imagine to define {\em internal observers} in the spirit of Ref. \cite{lorentzC}. 
In the latter reference, the authors postulate the existence of internal observers as observers living in a world of phonons,  
which can perform Michelson-Morley type interferometry with them. 
This gedankenexperiment then gives a zero fringe-shift (null) result explained by the Lorentz-FitzGerald contraction 
in such an internal phononic world.
However, the internal observer picture, in our context, comes with an inherent difficulty.
For nonlinear sound, signal propagation depends on the amplitude of perturbations [cf.~Eq.~\eqref{rw}) 
One would therefore be posed with the formidable task to construct, for obtaining a null result,  
an internal phonon interferometer which has nonlinear Lorentz-FitzGerald contraction for different signal amplitudes. 
On the other hand, even if general covariance is lost in the case of nonlinear sound, we can still define an internal observer in the phononic world, which is having access to the acoustic metric.  

\end{appendix}
\bibliography{nlgw_prd_v8}

\end{document}